\documentclass[a4paper,UKenglish,cleveref,autoref,thm-restate]{lipics-v2021}

\usepackage{microtype}

\usepackage{amsmath}
\usepackage{textcomp}
\usepackage{graphicx}
\usepackage{cite}

\usepackage{array}
\usepackage{booktabs}
\usepackage{multirow}
\usepackage{makecell}
\usepackage{tabularx}
\usepackage{threeparttable}

\usepackage[table]{xcolor}

\usepackage{listings}

\lstdefinestyle{javacode}{
    language=Java,
    basicstyle=\ttfamily\small,
    keywordstyle=\bfseries,
    commentstyle=\itshape\color{gray},
    stringstyle=\color{gray},
    numbers=left,
    numberstyle=\tiny\color{gray},
    numbersep=5pt,
    breaklines=true,
    frame=single,
    rulecolor=\color{black!30},
}
\usepackage{caption}
\usepackage{subcaption}

\usepackage{float}

\usepackage{multicol}

\usepackage{courier}

\usepackage{csquotes}
\usepackage{verbatim}
\usepackage{comment}

\usepackage{environ}
\usepackage{framed}
\usepackage{mdframed}
\usepackage[most]{tcolorbox}

\usepackage[noend]{algpseudocode}

\usepackage{tikz}

\usepackage{xurl}

\usepackage{soul}

\newcommand{\posgap}[1]{\cellcolor{red!25}#1}
\newcommand{\neggap}[1]{\cellcolor{blue!20}#1}
\newcommand{\zerogap}[1]{\cellcolor{gray!10}#1}

\definecolor{titleblue}{HTML}{1F5FAA}
\definecolor{codebg}{HTML}{FFFFFF}
\definecolor{codeborder}{HTML}{D0D0D0}
\definecolor{claimbg}{HTML}{F7F7F7}
\definecolor{failbg}{HTML}{FFF8E6}
\definecolor{failborder}{HTML}{D99A00}
\definecolor{kwcolor}{HTML}{005CC5}
\definecolor{strcolor}{HTML}{A31515}
\definecolor{commentcolor}{HTML}{22863A}
\definecolor{numcolor}{HTML}{6A737D}

\lstdefinestyle{lightcode}{
    language=Java,
    basicstyle=\ttfamily\fontsize{5.8}{6.4}\selectfont\color{black},
    keywordstyle=\bfseries\color{kwcolor},
    stringstyle=\color{strcolor},
    commentstyle=\itshape\color{commentcolor},
    numbers=left,
    numberstyle=\fontsize{5.5}{6}\selectfont\color{numcolor},
    numbersep=4pt,
    backgroundcolor=\color{codebg},
    frame=none,
    breaklines=true,
    breakatwhitespace=false,
    showstringspaces=false,
    tabsize=2,
    columns=fullflexible,
    keepspaces=true,
    xleftmargin=0.7em,
    aboveskip=0pt,
    belowskip=0pt
}

\newtcolorbox{examplepanel}[1]{
    enhanced,
    colback=white,
    colframe=gray!45,
    boxrule=0.5pt,
    arc=0pt,
    left=0pt,
    right=0pt,
    top=0pt,
    bottom=0pt,
    title=#1,
    coltitle=white,
    colbacktitle=titleblue,
    fonttitle=\bfseries\scriptsize,
    boxed title style={
        sharp corners,
        boxrule=0pt,
        colback=titleblue,
        left=5pt,
        right=5pt,
        top=3pt,
        bottom=3pt
    }
}

\newtcblisting{codebox}[1][]{
    listing only,
    listing options={style=lightcode},
    colback=codebg,
    colframe=codeborder,
    boxrule=0.4pt,
    arc=1pt,
    left=3mm,
    right=1mm,
    top=0.8mm,
    bottom=0.8mm,
    enhanced,
    before skip=2pt,
    after skip=2pt,
    #1
}

\newtcolorbox{claimbox}[1][]{
    colback=claimbg,
    colframe=claimbg,
    boxrule=0pt,
    arc=0pt,
    left=5pt,
    right=5pt,
    top=3pt,
    bottom=3pt,
    before skip=2pt,
    after skip=2pt,
    #1
}

\newtcolorbox{failurebox}[1][]{
    colback=failbg,
    colframe=failborder,
    boxrule=0.5pt,
    arc=1pt,
    left=5pt,
    right=5pt,
    top=3pt,
    bottom=3pt,
    before skip=2pt,
    after skip=0pt,
    #1
}

\renewcommand{\arraystretch}{1.1}
\newcolumntype{P}[1]{>{\centering\arraybackslash}m{#1}}

\newcounter{FindingsCounter}
\NewEnviron{finding}[1][]{\parindent0pt%
    \refstepcounter{FindingsCounter}
    \begin{tcolorbox}
        \textbf{Finding \arabic{FindingsCounter}:} \BODY
    \end{tcolorbox}
}

\newcounter{RQCounter}
\NewEnviron{researchquestion}[1][]{\parindent0pt%
    \refstepcounter{RQCounter}
    \begin{tcolorbox}
        $\mathbf{RQ_\arabic{RQCounter}}$: \BODY
    \end{tcolorbox}
    \vspace{0.1mm}
}

\title{Evaluating Language Models on Cross-Language Code Functional Equivalence}

\author{Hui Sun}
{North Carolina State University, Raleigh, NC, USA}
{hsun26@ncsu.edu}
{https://orcid.org/0009-0000-7911-3435}
{}

\author{Anderson Uchôa}
{Federal University of Ceará, Itapajé, Ceará, Brazil \and North Carolina State University, Raleigh, NC, USA}
{andersonuchoa@ufc.br}
{https://orcid.org/0000-0002-6847-5569}
{}

\author{Rohit Gheyi}
{Federal University of Campina Grande, Campina Grande, Paraíba, Brazil}
{rohit@dsc.ufcg.edu.br}
{https://orcid.org/0000-0002-5562-4449}
{}

\author{Wesley K. G. Assunção}
{North Carolina State University, Raleigh, NC, USA}
{wguezas@ncsu.edu}
{https://orcid.org/0000-0002-7557-9091}
{}

\authorrunning{H. Sun, A. Uchôa, R. Gheyi, and W.\,K.\,G. Assunção}

\Copyright{Hui Sun, Anderson Uchôa, Rohit Gheyi, and Wesley K. G. Assunção}



\category{Technical Track Paper}
\relatedversion{}
\supplementdetails[subcategory={Software, Data}]{Software}{https://doi.org/10.5281/zenodo.21800077}
\funding{This research was partially supported by FUNCAP (BP6-00241-00276.01.00/25), CNPq (201708/2025-6, 403304/2025-3, 404406/2023-8, 403719/2024-0), FAPESQ-PB (268/2025), CAPES, and Google Cloud Research Credits Program with the award GCP19980904.}

\EventEditors{Robert Feldt, Maria Paasivaara, Daniel Mendez, Stefan Wagner, and Marvin Mu\~{n}oz Bar\'{o}n}
\EventNoEds{5}
\EventLongTitle{20th International Symposium on Empirical Software Engineering and Measurement (ESEM 2026)}
\EventShortTitle{ESEM 2026}
\EventAcronym{ESEM}
\EventYear{2026}
\EventDate{October 8--9, 2026}
\EventLocation{Munich, Germany}
\EventLogo{}
\SeriesVolume{394}
\ArticleNo{41}

\nolinenumbers

\begin{document}

\maketitle

\begin{abstract}
\textbf{Background.} Large Language Models (LLMs) have demonstrated strong performance across a variety of code-understanding tasks, leading many to believe that they can reason about program semantics. A fundamental test of this capability is assessing code functional equivalence. However, existing evaluations primarily focus on single-language settings or rely on synthetically generated code, raising concerns about whether current results reflect true semantic understanding.
\noindent\textbf{Aims.} We investigate whether LLMs can accurately judge functional equivalence across different programming languages in human-written code, a setting that requires deeper reasoning beyond superficial similarity.
\noindent\textbf{Method.} We introduce PolyHuman, a dataset of human-written programs in CPP, Java, and Python. Using this dataset, we evaluate intra- and inter-language equivalence detection across open-weight and proprietary LLMs, selecting GPT-o4-mini as a representative model to assess stability. We then manually analyze 81 cases of systematic disagreement in which models incorrectly judge functional equivalence, examining the code logic and the generated Chain-of-Thought reasoning. Finally, we categorize these failures and compare them across GPT-o4-mini, Claude-Opus-4.7, and Gemini-3-Flash to determine whether they reflect model-specific issues or broader limitations of state-of-the-art LLMs.
\noindent\textbf{Results.} We identify a difficulty-dependent breakdown in equivalence judgment (harder problems make the model increasingly prone to misclassifying non-equivalent code as equivalent), a model-specific sensitivity to programming language for the best-performing model (particularly a more conservative behavior on Python), and a partial reliance on similarity-based cues. GPT-o4-mini also shows substantial run-to-run instability under identical settings, indicating inconsistent rather than absent capability. Its failures are mainly due to (1) a lack of knowledge relevant to judging code functional equivalence and (2) abstraction-level reasoning failure, such as premature conclusion, semantic-level misunderstanding, and implementation-level error. We also observed that several analyzed failures involved loops and conditional branches.
\noindent\textbf{Conclusions.} Current LLMs do not reliably capture functional equivalence within or across languages. Our findings highlight the need for more realistic benchmarks and suggest improvements through better reasoning-path selection and by addressing reasoning gaps across the identified failure levels.
\end{abstract}

\section{Introduction}


Large Language Models (LLMs) have recently demonstrated remarkable progress in code understanding tasks, including generation, summarization, and bug fixing~\cite{jiang2026survey, wang2021codet5, VanshikaVolvo2025, PankajForge2026}. These advances have led to growing expectations that LLMs can reason about code semantics at a level comparable to human developers~\cite{ahmad2021unified, wang2021codet5, NafisaMSR2025}. Another fundamental capability underlying such reasoning is assessing code functional equivalence, which is the ability to determine whether two programs produce identical outputs for all valid inputs~\cite{maveli2025can, puri2021codenet, wei2025equibench}. This capability is critical for applications such as code refactoring~\cite{fowler2018refactoring}, cross-language translation~\cite{pan2024lost}, code migration~\cite{nikolov2026multi, AssuncaoTOSEM2025}, plagiarism detection~\cite{prechelt2002finding, schleimer2003winnowing}, and compiler optimization~\cite{maruthamuthu2023advancements, muchnick1997advanced}.

Despite strong reported performance, we argue that current evaluations overestimate LLMs' true semantic understanding of code. Existing benchmarks for functional equivalence largely focus on intra-language settings~\cite{gheyi2026oracle,saferefactor-ieee,svajlenko2014towards} or rely on synthetically transformed programs~\cite{rolim2017learning, yu2019neural}, where equivalence can often be inferred from superficial cues such as lexical similarity or minor structural changes~\cite{roy2007survey,wei2025equibench}. These evaluations fail to capture the complexity of independently written human programs. 
This limitation becomes pronounced in cross-language equivalence detection (e.g., after code translation~\cite{pan2024lost}), where semantically identical programs may differ significantly in syntax, structure, and idiomatic patterns~\cite{moumoula2025struggles, perez2019cross}. In such settings, shallow similarity signals break down, requiring deeper reasoning about program behavior~\cite{moumoula2025struggles, wei2025equibench}. However, it remains unclear whether LLMs possess this capability or whether their apparent success is driven by dataset artifacts and heuristic shortcuts.

To address this gap, we conduct the first systematic investigation of LLMs’ ability to judge functional equivalence across programming languages using human-written code. We introduce PolyHuman, a new benchmark constructed from competitive programming platforms, containing program pairs in CPP (i.e., C++), Java, and Python. Unlike prior datasets~\cite{rolim2017learning, svajlenko2014towards, yu2019neural}, PolyHuman captures realistic implementation variation, enabling a more rigorous evaluation of semantic reasoning.
Using PolyHuman, we examine how well LLMs perform when comparing programs written within the same language versus across different programming languages, and whether their performance degrades in the presence of human-written implementations rather than synthetic transformations. We further analyze the factors that influence model predictions, including code complexity, similarity signals, and language-specific characteristics, and seek to uncover the underlying causes of model failures, particularly in cross-language scenarios where semantic reasoning is most critical.

Guided by these goals, our study addresses four research questions. RQ$_1$ asks \textit{how LLMs assess functional equivalence in intra-language settings}, establishing a baseline for model performance. RQ$_2$ extends this to \textit{inter-language settings}, examining accuracy degradation relative to the intra-language baseline (e.g., for language pairs such as CPP$\leftrightarrow$Java or Python$\leftrightarrow$Java) and the extent to which models generalize beyond language-specific syntax, thereby probing whether LLMs rely on genuine semantic reasoning or superficial cues. RQ$_3$ investigates \textit{what factors influence LLM assessment of functional equivalence}, including the relationship between lexical and structural similarity and prediction outcomes, the impact of code complexity, and models' sensitivity to specific programming languages, particularly in cross-language comparisons. Finally, RQ$_4$ examines \textit{what causes LLM failure} in functional equivalence assessment, through a detailed error analysis of incorrect predictions that reveals systematic biases (e.g., overprediction of non-equivalence) and recurring error patterns.

Our results challenge the assumption that LLMs possess robust semantic code understanding, showing that:
(1) LLM performance is inflated by datasets with balanced labels and syntactic transformations~\cite{maveli2025can, wei2025equibench};
(2) models exhibit systematic and model-dependent prediction biases, some skewing toward conservative “non-equivalent” judgments and others toward over-accepting equivalence, an effect most pronounced in cross-language settings;
(3)~human-written programs introduce substantial difficulty, exposing a significant gap between synthetic and real-world evaluation scenarios; and
(4) the best-performing model's decisions are partially influenced by similarity signals, indicating incomplete and inconsistent reasoning mechanisms. 
Overall, our findings suggest that current LLMs do not reliably reason about functional equivalence, especially across languages, and that new evaluation paradigms are necessary to accurately assess semantic code understanding. 

This paper makes four main contributions. First, we introduce PolyHuman, a dataset of human-written CPP, Java, and Python programs for evaluating functional equivalence within and across languages. Second, we conduct the first large-scale empirical study of LLMs on cross-language functional equivalence assessment. Third, to the best of authors' knowledge, we are the first to characterize LLM functional equivalence judgment and provide a systematic codebook revealing the capability gaps at each level. Finally, we derive a set of actionable implications for LLM researchers, agent system designers, and programming practitioners, offering concrete guidance on model selection, human oversight, and risk-aware deployment in code-related tasks.


\section{Methodology}

To evaluate whether LLMs can accurately judge functional equivalence, we design an empirical study over human-written programs in two settings: (1) \textit{intra-language}, where both programs are written in the same programming language; and (2) \textit{inter-language}, where they are written in different languages.
To systematically investigate this problem, we adopt the Goal-Question-Metric (GQM) framework~\cite{caldiera1994goal}. 
Our \textbf{\textit{goal}} is to analyze the ability of LLMs to detect functional equivalence; 
\textbf{\textit{for the purpose of}} assessing whether their performance reflects true semantic reasoning; 
\textbf{\textit{with respect to}} accuracy across settings, performance degradation on human-written code, the influence of similarity and language, and the causes of models' failure; 
\textbf{\textit{from the perspective of}} researchers and developers; 
\textbf{\textit{in the context of}} human-written programs in CPP, Java, and Python. 

\subsection{PolyHuman Dataset Construction}\label{datasetConstruction}

To enable a realistic evaluation of functional equivalence detection, we construct PolyHuman, a dataset of human-written programs across programming languages, as detailed below. 

\noindent\textbf{Data Source.}  To construct PolyHuman, we build upon CodeContests~\cite{li2022competition}, a large-scale dataset containing correct and incorrect human-written solutions to 13,610 programming problems, collected from five competitive programming platforms (e.g., Aizu, AtCoder, Codeforces). Each problem in CodeContests is annotated with a difficulty rating, from \textit{A (easy)} to \textit{G (expert level)}, derived from the original platform (e.g., Codeforces rating system). 
We chose CodeContests because it includes realistic characteristics related to code length variation, language coverage (CPP, Java, and Python), and solution correctness. These characteristics enable us to analyze how problem complexity and domain influence LLM performance in functional equivalence assessment.

\noindent\textbf{Filtering Strategy.} Online judges report various submission outcomes, including wrong answer, time limit exceeded, memory limit exceeded, runtime error, and presentation error. In this study, we consider a solution to be \textit{correct} if it passes all test cases, and \textit{incorrect} otherwise. Under this definition, when evaluating code pairs, any combination of a ``correct'' and an ``incorrect'' solution inherently exhibits at least one point of behavioral divergence on the test suite. This encapsulation allows us to focus strictly on observable input-output behavior rather than specific failure causes, thereby operationally labeling such pairs as non-equivalent. We retain only CodeContests' solutions in three popular programming languages, namely CPP, Java, and Python. This filtering ensures the construction of both equivalent and non-equivalent pairs across and within languages.

\noindent\textbf{Sampling and Randomization.} For each selected programming problem, CodeContests contains multiple solutions for each language. 
To avoid bias toward specific implementations or coding styles, we adopt a randomized sampling strategy. Specifically, for each problem and programming language (CPP, Java, and Python), we randomly sample two correct (passing) solutions and one incorrect (failing) solution. These sampled solutions are then combined to construct intra-language and inter-language pairs. This process ensures diversity in implementations while preventing over-representation of any particular solution pattern.

\noindent\textbf{Sub-tasks and Code Pair.} From the filtered and sampled solutions, we construct four categories of code pairs, totaling 15 s. We treat pairs of correct solutions as functionally equivalent, and pairs containing one incorrect solution as non-equivalent. After filtering, we obtain 6,905 problems. Excluding problems with unknown difficulty yields the final PolyHuman dataset comprising 5,035 problems.
Table~\ref{tab:pair_categories} summarizes the pair categories.
%






%

\begin{table}[ht!]
\centering
\caption{Categories of code pairs used to construct the s.}
\label{tab:pair_categories}
\scriptsize
\setlength{\tabcolsep}{4pt}
\renewcommand{\arraystretch}{1.12}
\resizebox{\linewidth}{!}{ 
\begin{tabular}{p{3cm} p{2cm} >{\raggedright\arraybackslash}p{7.5cm}}
\toprule
\textbf{Category} & \textbf{Equivalence} & \textbf{Examples} \\
\midrule

Intra-language
& Equivalent 
& JAVA\_Pass1 vs. JAVA\_Pass2; CPP\_Pass1 vs. CPP\_Pass2; PYTHON3\_Pass1 vs. PYTHON3\_Pass2 \\ \midrule

Intra-language 
& Non-equivalent 
& JAVA\_Pass1 vs. JAVA\_Fail; CPP\_Pass1 vs. CPP\_Fail; PYTHON3\_Pass1 vs. PYTHON3\_Fail \\\midrule

Inter-language
& Equivalent 
& JAVA\_Pass1 vs. PYTHON3\_Pass1; JAVA\_Pass1 vs. CPP\_Pass1; CPP\_Pass1 vs. PYTHON3\_Pass1 \\\midrule

Inter-language
& Non-equivalent 
& JAVA\_Fail vs. CPP\_Pass1; JAVA\_Fail vs. PYTHON3\_Pass1; CPP\_Fail vs. JAVA\_Pass1; CPP\_Fail vs. PYTHON3\_Pass1; PYTHON3\_Fail vs. JAVA\_Pass1; PYTHON3\_Fail vs. CPP\_Pass1 \\

\bottomrule
\end{tabular}
}
\end{table}

\subsection{Experiment Setup}\label{experimentDesign}

This section describes the datasets, models, and experimental settings used in our evaluation.

\noindent\textbf{Baseline Benchmarks.}  
We consider two widely used benchmarks for code functional equivalence: EquiBench~\cite{wei2025equibench} and SeqCoBench~\cite{maveli2025can}. 
EquiBench consists of 2,400 program pairs across four languages and six categories (\textit{DCE}, \textit{CUDA}, \textit{x86-64}, \textit{OJ-A}, \textit{OJ-V}, and \textit{OJ-VA}). In this study, we focus on \textit{OJ-A} and \textit{OJ-V}, which contain Python programs derived from competitive programming platforms. The distinction between them lies in their construction: \textit{OJ-A} contains pairs of independently written human programs, whereas \textit{OJ-V} includes pairs generated through variable renaming transformations~\cite{flook2025python}. Each subset includes 200 equivalent and 200 non-equivalent pairs. 
SeqCoBench is a Python-based benchmark composed of code pairs generated through 20 types of semantic-preserving (e.g., Function/identifier renaming and
Literal/value substitutions) and semantic-altering transformations (e.g., logic-changing operator flips and subtle semantic perturbations intended to preserve lexical similarity). 

Since our work relies on running language models, which entail computational costs (for open-weight models) and monetary costs (for proprietary models), we evaluate on a subset of instances from the PolyHuman and SeqCoBench datasets. For SeqCoBench, we randomly sampled 500 instances. For PolyHuman, we used the first 500 instances as ordered in the dataset, which is not sorted by difficulty, model performance, or any other property. For EquiBench, we used all 800 instances (400 OJ-A + 400 OJ-V). Table~\ref{tab:dataset-stats} summarizes key statistics for EquiBench, SeqCoBench, and PolyHuman instances used in our study.

\begin{table}[ht!]
\centering
\scriptsize
\setlength{\tabcolsep}{3pt}
\renewcommand{\arraystretch}{1.1}
\caption{Dataset statistics comparison across EquiBench, SeqCoBench, and PolyHuman.}
\label{tab:dataset-stats}
\begin{tabular}{llcccccc}
\toprule
\multirow{2}{*}{\textbf{Category}} & \multirow{2}{*}{\textbf{Metric}}
& \multicolumn{2}{c}{\textbf{EquiBench}}
& \textbf{SeqCoBench}
& \multicolumn{3}{c}{\textbf{PolyHuman}} \\
\cmidrule(lr){3-4} \cmidrule(lr){5-5} \cmidrule(lr){6-8}
& & OJ-A & OJ-V & Sampled Subset & \multicolumn{3}{c}{Each  (15 total)} \\
\midrule

Language & & Python & Python & Python & CPP & Java & Python \\ \midrule

\multirow{2}{*}{Pairs}
& Equal     & 200 & 200 & 250 & 500 & 500 & 500 \\
& Not Equal & 200 & 200 & 250 & 500 & 500& 500 \\ \midrule

\multirow{3}{*}{Lines of Code}
& Min  & 3    & 2    & 2    & 1     & 1      & 1 \\
& Max  & 3,403 & 4,087 & 50   & 1,233  & 1,423   & 1,042 \\
& Avg. & 82   & 70   & 7.79 & 48.52 & 109.88 & 34.78 \\
\bottomrule
\end{tabular}
\end{table}

EquiBench (\textit{OJ-A} and \textit{OJ-V}) and SeqCoBench are balanced, Python-only benchmarks. Each EquiBench subset contains 200 pairs per label, with average LOC of 82 and 70, respectively. From SeqCoBench's balanced pool of 1,860 pairs per label, we sample 250 per label (average LOC: 7.79 in Table~\ref{tab:dataset-stats}). PolyHuman spans CPP, Java, and Python through 15 label-homogeneous sub-tasks, each containing 500 pairs from the same 500 sampled problems (of 5,035 total); six sub-tasks are equivalent and nine non-equivalent (Table~\ref{tab:pair_categories}). Average LOC is 48.52, 109.88, and 34.78 for CPP, Java, and Python, respectively, consistent with known differences in language verbosity~\cite{prechelt2002empirical}.

\noindent\textbf{Evaluated LLMs.} 
We evaluate nine language models shown in Table~\ref{tab:model_performance_1}. These models were selected to cover both proprietary and open-source architectures with varying sizes and capabilities, enabling a comprehensive comparison of performance across different model families.
All experiments are conducted using deterministic decoding (temperature = 0) to ensure reproducibility, except for GPT-o4-mini, whose temperature cannot be set.



\noindent\textbf{Prompting Strategy.} We adopt a consistent prompting strategy across all models and datasets to ensure fair comparison. For each code pair, a single prompt is issued to the model, containing the two programs to be evaluated.
In both settings (i.e., \textit{intra-language} and \textit{inter-language}), the task remains identical; the model is asked to determine whether the two programs are functionally equivalent. The same prompt template is applied across EquiBench and PolyHuman; SeqCoBench does not include a problem description.

The prompt follows a zero-shot format and explicitly instructs the model to reason about program behavior rather than relying on superficial similarity. To reduce ambiguity and enforce consistent outputs, we constrain model responses to a binary decision (\textit{Yes} or \textit{No}). 

\noindent\textbf{Error Analysis.} To understand model failure modes, we conduct both quantitative and qualitative analyses of incorrect predictions. We examine error distributions (e.g., false positives vs. false negatives), analyze error rates across programming languages, and identify recurring patterns in misclassified pairs. Furthermore, we conduct a fine-grained comparative analysis on a curated set of challenging instances (i.e., those on which GPT-o4-mini repeatedly produces incorrect predictions). In these instances, we compare the behavior of GPT-o4-mini, Gemini-3-Flash, and Claude-Opus-4.7 to characterize where and why models diverge, uncovering model-specific failure modes such as premature conclusions and knowledge gaps.

\section{Results and Discussion}

\subsection{LLM Performance for Same Programming Language Pairs (\texorpdfstring{RQ$_1$}{RQ1})}

\textbf{RQ$_1$} compares EquiBench~\cite{wei2025equibench} (\textit{OJ-A} and \textit{OJ-V}), SeqCoBench~\cite{maveli2025can}, and PolyHuman intra-language pairs using the common models, prompt, and accuracy metric described in Section~\ref{experimentDesign}.
This unified evaluation bridges gaps in prior studies: EquiBench does not include code-specialized LLMs, while SeqCoBench does not evaluate closed-source, proprietary models. By standardizing both models and metrics, we enable a more consistent and comprehensive comparison across benchmarks.
We additionally inspect label-specific prediction rates to identify systematic decision biases.
Table~\ref{tab:model_performance_1} shows the LLMs' accuracy on the baselines (EquiBench and SeqCoBench) and intra-language instances from PolyHuman.

\begin{table}[ht!]
\scriptsize
\centering
\caption{Model accuracy across baseline datasets and PolyHuman intra-language tasks.}
\label{tab:model_performance_1}
\resizebox{\linewidth}{!}{ 
\begin{tabular}{l|ccc|ccc|ccc}
\toprule
\multirow{2}{*}{\textbf{Model}} 
& \multicolumn{3}{c|}{\textbf{Baselines}} 
& \multicolumn{6}{c}{\textbf{PolyHuman (Intra-language)}} \\

\cmidrule(lr){2-4} \cmidrule(lr){5-10}

& \multicolumn{2}{c}{EquiBench} & \multirow{2}{*}{SeqCoBench} 
& \multicolumn{3}{c}{\textbf{Pass vs Pass}} 
& \multicolumn{3}{c}{\textbf{Pass vs Fail}} \\

\cmidrule(lr){5-7} \cmidrule(lr){8-10}

& OJ-V & OJ-A & 
& CPP & Java & Python 
& CPP & Java & Python \\

\midrule

GPT-o4-mini               & 0.91 & 0.82 & 0.97 & 0.83 & 0.84 & 0.80 & 0.83 & 0.89 & 0.93 \\
GPT-OSS-20B               & 0.89 & 0.74 & 0.95 & 0.59 & 0.62 & 0.66 & 0.91 & 0.94 & 0.93 \\
Llama-3.1-8B-Instruct     & 0.54 & 0.54 & 0.84 & 0.98 & 0.98 & 0.98 & 0.05 & 0.04 & 0.07 \\
Llama-3.2-3B-Instruct     & 0.52 & 0.53 & 0.66 & 0.54 & 0.44 & 0.78 & 0.53 & 0.66 & 0.28 \\
Mistral-7B-Instruct-v0.3  & 0.51 & 0.51 & 0.50 & 1.00 & 1.00 & 1.00 & 0.03 & 0.02 & 0.02 \\
Qwen2.5-Coder-7B-Instruct & 0.67 & 0.64 & 0.89 & 0.66 & 0.75 & 0.66 & 0.54 & 0.40 & 0.54 \\
Qwen2.5-Coder-14B-Instruct & 0.74  & 0.71 & 0.87 & 0.60 & 0.56 & 0.50 & 0.65 & 0.66 & 0.71 \\
CodeLlama-7B              & 0.57 & 0.54 & 0.77 & 0.98 & 0.99 & 0.96 & 0.03 & 0.01 & 0.06 \\
CodeLlama-13B              & 0.57 & 0.54 & 0.75 & 0.31 & 0.53 & 0.30 & 0.74 & 0.52 & 0.76 \\
\midrule
\textbf{Average}          & 0.66 & 0.62 & 0.80 & 0.72 & 0.75 & 0.74 & 0.48 & 0.46 & 0.48\\
\bottomrule

\end{tabular}
}
\scriptsize{\textit{Note: Each cell reports the accuracy of pairs in the same programming language.
}}
\vspace{-0.3cm}
\end{table}

\noindent\textbf{Bias-Induced Accuracy Illusion in Models.} 
We observe that several LLMs exhibit strong label prediction bias in code functional equivalence tasks. Although their accuracy on EquiBench~\cite{wei2025equibench} is often close to 50\%, this does not indicate random behavior. Instead, models tend to collapse into near-constant predictions (e.g., predominantly predicting \textit{Yes} or \textit{No}). Because EquiBench is balanced (50/50), such biased behavior can produce accuracy close to random guessing, creating an illusion of reasonable performance.
Using PolyHuman, we uncover clear and systematic biases across models. For instance, Llama-based models and Qwen2.5-Coder tend to predict \textit{non-equivalence}, whereas Mistral-7B-Instruct-v0.3 shows a tendency toward \textit{equivalence}. This bias persists even in stronger models. For example, specifically in Python, GPT-o4-mini achieves only 0.80 accuracy on \textit{Python Pass vs. Pass} pairs (where both code snippets are correct and equivalent), but 0.93 on \textit{Python Pass vs. Fail} (where one snippet is correct and the other is faulty/non-equivalent). This discrepancy indicates that the model is biased toward predicting non-equivalence, particularly within the Python language. This bias is particularly pronounced for GPT-OSS-20B across all three programming languages.

\begin{tcolorbox}[enhanced, boxsep=0pt, before skip=5pt, after skip=5pt]
\textbf{Finding 1:} Across the studied LLMs, each model exhibits its own systematic bias in equivalence judgments, consistently skewing toward either false positives or false negatives rather than erring symmetrically.\end{tcolorbox}

\noindent\textbf{Human-written code equivalence tasks were more challenging for GPT-o4-mini.} Using GPT-o4-mini as a representative baseline~\cite{wei2025equibench}, we observe a consistent performance decline as evaluation benchmarks shift from synthetic to human-written programs, reflecting the impact of increasing code realism. The model achieves near-perfect accuracy on synthetic datasets like SeqCoBench ($0.97$; Avg. LOC: $7.79$) and high accuracy on OJ-V ($0.91$; Avg. LOC: $70$), where equivalence is derived from simple variable renaming. 

In contrast, performance decreases substantially on benchmarks with independently written implementations: OJ-A ($0.82$; Avg. LOC: $82$) and PolyHuman's Python subset ($0.87$; Avg. LOC: $34.78$), kept Python-only here for a like-for-like comparison with the other Python-based baselines. Furthermore, the performance on PolyHuman's Python pairs reveals a distinct systematic bias: while the model reliably detects non-equivalence (\textit{Pass vs. Fail}, $0.93$ accuracy), it struggles to recognize equivalence between structurally diverse implementations (\textit{Pass vs. Pass}, $0.80$ accuracy).

\begin{tcolorbox}[enhanced, boxsep=0pt, before skip=5pt, after skip=0pt]
\textbf{Finding 2:}
For GPT-o4-mini, accuracy is high on synthetic datasets but drops sharply on human-written code, suggesting that independently written implementations introduce greater structural diversity, making equivalence detection more difficult. GPT-o4-mini and GPT-OSS-20B also tend to detect non-equivalent code more reliably than truly equivalent solutions with different implementations; averaged across all evaluated models, however, the opposite pattern holds (Pass-Pass $\approx 0.72$--$0.75$ vs. Pass-Fail $\approx 0.46$--$0.48$; Table~\ref{tab:model_performance_1}), reflecting the over-acceptance bias exhibited by several other models.
\end{tcolorbox}

\subsection{LLM Performance for Different Programming Language Pairs (\texorpdfstring{RQ$_2$}{RQ2})}

To answer \textbf{RQ$_2$}, we analyze how LLM performance varies when models judge functional equivalence across different programming languages. While in RQ$_1$ we focus on intra-language comparisons, in RQ$_2$ we evaluate inter-language pairs from PolyHuman, covering both equivalent and non-equivalent cases. With this analysis, we quantify performance differences across language combinations and examine whether specific languages or pair types are associated with systematic prediction biases.
Table~\ref{tab:functional_equivalence_accuracy} reports the LLMs' accuracy across inter-language pairs. The first group of columns represents equivalent pairs, while the remaining columns represent non-equivalent pairs. In the latter, ``F'' marks the language associated with the failing case.

\begin{table*}[ht!]
\centering
\renewcommand{\arraystretch}{1.15}
\caption{Accuracy of functional equivalence detection across inter-language code pairs.}
\label{tab:functional_equivalence_accuracy}
\scriptsize
\resizebox{\textwidth}{!}{%
\addtolength{\tabcolsep}{-2pt}
\begin{tabular}{lccc|cccccc}
\toprule
\multirow{2}{*}{\textbf{Model}} 
& \multicolumn{3}{c|}{\textbf{Equivalent pairs (Pass--Pass)}} 
& \multicolumn{6}{c}{\textbf{Non-equivalent pairs (Pass--Fail)}} \\
\cmidrule(lr){2-4}
\cmidrule(lr){5-10}
& \makecell{\textbf{CPP}\\$\leftrightarrow$\\\textbf{Java}} 
& \makecell{\textbf{Java}\\$\leftrightarrow$\\\textbf{Python}} 
& \makecell{\textbf{Python}\\$\leftrightarrow$\\\textbf{CPP}} 
& \makecell{\textbf{Java(F)}\\$\leftrightarrow$\\\textbf{CPP}} 
& \makecell{\textbf{Java(F)}\\$\leftrightarrow$\\\textbf{Python}} 
& \makecell{\textbf{CPP(F)}\\$\leftrightarrow$\\\textbf{Java}} 
& \makecell{\textbf{CPP(F)}\\$\leftrightarrow$\\\textbf{Python}} 
& \makecell{\textbf{Python(F)}\\$\leftrightarrow$\\\textbf{Java}}
& \makecell{\textbf{Python(F)}\\$\leftrightarrow$\\\textbf{CPP}}
\\
\midrule
GPT-o4-mini 
& 0.85 & 0.82 & 0.81 
& 0.93 & 0.90 & 0.89 & 0.87 & 0.92 & 0.92 \\

GPT-OSS-20B 
& 0.63 & 0.62 & 0.64 
& 0.92 & 0.91 & 0.88 & 0.86 & 0.90 & 0.92 \\

Llama-3.1-8B-Instruct 
& 1.00 & 0.99 & 0.99 
& 0.02 & 0.03 & 0.02 & 0.03 & 0.02 & 0.02 \\

Llama-3.2-3B-Instruct 
& 0.51 & 0.61 & 0.72 
& 0.56 & 0.44 & 0.54 & 0.35 & 0.53 & 0.31 \\

Mistral-7B-Instruct-v0.3 
& 1.00 & 1.00 & 1.00 
& 0.01 & 0.01 & 0.01 & 0.01 & 0.01 & 0.02 \\

Qwen2.5-Coder-7B-Instruct 
& 0.80 & 0.63 & 0.67 
& 0.38 & 0.47 & 0.31 & 0.48 & 0.30 & 0.43 \\

Qwen2.5-Coder-14B-Instruct 
& 0.62 & 0.55 & 0.53 
& 0.55 & 0.60 & 0.58 & 0.62 & 0.63 & 0.60 \\

CodeLlama-7B 
& 0.99 & 0.97 & 0.98 
& 0.03 & 0.04 & 0.01 & 0.05 & 0.02 & 0.02 \\

CodeLlama-13B 
& 0.48 & 0.30 & 0.22 
& 0.67 & 0.73 & 0.59 & 0.81 & 0.64 & 0.78 \\
\bottomrule
\end{tabular}%
}
\end{table*}

The results show that model behavior and performance vary across language pairs. To characterize these differences, we computed a decision gap for each LLM and programming language as $Gap = AccNonEq - AccEq$, where $AccEq$ and $AccNonEq$ represent the average accuracies on equivalent and non-equivalent pairs involving that language, respectively. A positive gap indicates a tendency to over-reject equivalence, whereas a negative gap suggests a tendency to over-accept equivalence. 
Table~\ref{tab:model_language_bias_gap} presents the bias gaps, with colors to highlight the direction: \colorbox{red!25}{red} indicates a bias toward non-equivalence (positive gap), \colorbox{blue!20}{blue} indicates a bias toward equivalence (negative gap), and \colorbox{gray!10}{gray} represents near-balanced behavior.

\begin{table*}[ht!]
\centering
\renewcommand{\arraystretch}{1.15}
\setlength{\tabcolsep}{4pt}
\caption{Model-level decision gap across equivalent and non-equivalent inter-language pairs.}
\label{tab:model_language_bias_gap}
\scriptsize
\resizebox{\textwidth}{!}{%
\begin{tabular}{lccc ccc ccc}
\toprule
\textbf{Model} 
& \multicolumn{3}{c}{\textbf{CPP}} 
& \multicolumn{3}{c}{\textbf{Java}} 
& \multicolumn{3}{c}{\textbf{Python}} \\
\cmidrule(lr){2-4}
\cmidrule(lr){5-7}
\cmidrule(lr){8-10}
& \textbf{AccEq} & \textbf{AccNonEq} & \textbf{Gap}
& \textbf{AccEq} & \textbf{AccNonEq} & \textbf{Gap}
& \textbf{AccEq} & \textbf{AccNonEq} & \textbf{Gap} \\
\midrule
GPT-o4-mini 
& 0.830 & 0.902 & \posgap{0.072}
& 0.835 & 0.910 & \posgap{0.075}
& 0.815 & 0.902 & \posgap{0.088} \\

GPT-OSS-20B 
& 0.635 & 0.895 & \posgap{0.260}
& 0.625 & 0.902 & \posgap{0.277}
& 0.630 & 0.898 & \posgap{0.267} \\

Llama-3.1-8B-Instruct 
& 0.995 & 0.023 & \neggap{-0.972}
& 0.995 & 0.023 & \neggap{-0.972}
& 0.990 & 0.025 & \neggap{-0.965} \\

Llama-3.2-3B-Instruct 
& 0.615 & 0.440 & \neggap{-0.175}
& 0.560 & 0.518 & \neggap{-0.042}
& 0.665 & 0.408 & \neggap{-0.258} \\

Mistral-7B-Instruct-v0.3 
& 1.000 & 0.012 & \neggap{-0.988}
& 1.000 & 0.010 & \neggap{-0.990}
& 1.000 & 0.012 & \neggap{-0.988} \\

Qwen2.5-Coder-7B-Instruct 
& 0.735 & 0.400 & \neggap{-0.335}
& 0.715 & 0.365 & \neggap{-0.350}
& 0.650 & 0.420 & \neggap{-0.230} \\

Qwen2.5-Coder-14B-Instruct 
& 0.575 & 0.588 & \zerogap{0.013}
& 0.585 & 0.590 & \zerogap{0.005}
& 0.540 & 0.612 & \posgap{0.073} \\

CodeLlama-7B 
& 0.985 & 0.028 & \neggap{-0.958}
& 0.980 & 0.025 & \neggap{-0.955}
& 0.975 & 0.032 & \neggap{-0.942} \\

CodeLlama-13B 
& 0.350 & 0.713 & \posgap{0.363}
& 0.390 & 0.658 & \posgap{0.267}
& 0.260 & 0.740 & \posgap{0.480} \\
\bottomrule
\end{tabular}%
}
\end{table*}

We observe three distinct decision profiles based on the performance gap between equivalent and non-equivalent pairs. 
First, \textbf{over-rejection of equivalence} is exhibited by GPT-o4-mini, GPT-OSS-20B, Qwen2.5-Coder-14B-Instruct, and CodeLlama-13B, all showing positive gaps across languages. This bias is strongest in CodeLlama-13B, particularly for Python pairs ($Gap=0.480$, $AccEq=0.260$, $AccNonEq=0.740$). Second, \textbf{extreme over-acceptance} is found in Llama-3.1-8B-Instruct, Mistral-7B-Instruct-v0.3, and CodeLlama-7B, which present negative gaps. Notably, Mistral-7B-Instruct-v0.3 achieves $AccEq=1.000$ across all languages but fails entirely on non-equivalent pairs. Third, \textbf{mild over-acceptance} is observed in Llama-3.2-3B-Instruct and Qwen2.5-Coder-7B-Instruct, which display intermediate negative gaps.

Among the evaluated models, Qwen2.5-Coder-14B-Instruct shows the most balanced gap. Its gaps remain close to zero for CPP ($Gap=0.013$) and Java ($Gap=0.005$), with only a small positive gap for Python ($Gap=0.073$). This does not imply the highest absolute accuracy, but it suggests lower decision bias compared with models showing extreme positive or negative gaps. GPT-o4-mini is also competitive due to its relatively high accuracy on both equivalent and non-equivalent pairs, although its consistently positive gaps indicate a more conservative profile.

Additionally, Wilcoxon signed-rank tests~\cite{whitley2002statistics} comparing $AccEq$ and $AccNonEq$ within each language find no statistically significant difference (CPP: $W=13.0$, $p=0.30$; Java: $W=13.0$, $p=0.30$; Python: $W=14.0$, $p=0.36$; $n=9$). A Friedman test~\cite{friedman1937use} comparing bias gaps ($Gap=AccNonEq-AccEq$) across languages is borderline ($\chi^2=5.35$, $df=2$, $p=0.069$, Kendall's $W=0.297$), suggesting a possible but statistically inconclusive language effect. We therefore do not claim that language plays no role; rather, CPP, Java, and Python performance consistently track each model's own decision tendency more closely than any common language pattern, and the wide variation in gap magnitude across models (from $-0.988$ for Mistral-7B-Instruct-v0.3 to $+0.363$ for CodeLlama-13B in CPP) points to model identity as the more salient factor. This aggregate, cross-model test ($n=9$) has limited power and does not rule out the model-specific Python effect found in RQ$_3$'s larger-sample regression ($\beta=-0.450$, $p=0.001$; Table~\ref{tab:regression_results}); language thus appears secondary and model-specific rather than a uniformly shared driver.

Practically, these results show that model-specific differences were larger than the common language effect in our evaluated set.
For instance, Mistral-7B-Instruct-v0.3, Llama-3.1-8B-Instruct, and CodeLlama-7B exhibit high false-equivalence rates, whereas CodeLlama-13B over-rejects equivalence. To mitigate these complementary biases, ensemble-based judging strategies offer a promising solution~\cite{zhou2025se}. This aligns with recent findings that LLM capabilities vary significantly across programming contexts and tasks~\cite{feng2024complexcodeeval,izadi2024language,yang2024evaluation,yu2024codereval}.

\begin{tcolorbox} [enhanced, boxsep=0pt, before skip=5pt, after skip=0pt]
\textbf{Finding 3:}
The studied LLMs do not behave equally in cross-language equivalence judgments: some tend to over-reject equivalence, while others tend to over-accept it. Among the models evaluated, Qwen2.5-Coder-14B-Instruct shows the most balanced behavior, while GPT-o4-mini offers the best overall trade-off. These results show that model choice matters: highly biased models should not be used alone, and combining complementary models may lead to more reliable judgments.

\end{tcolorbox}

\subsection{Factors Influencing the Best-Performing Model's Predictions (\texorpdfstring{RQ$_3$}{RQ3})}

To answer \textbf{RQ$_3$}, we conduct an in-depth behavioral analysis of GPT-o4-mini (the top-performing model in RQ$_2$) to evaluate how code- and task-level features from PolyHuman influence its functional equivalence judgments. Specifically, we extract three categories of features: (1) problem difficulty (levels A to G); (2) code length (LOC); and (3) superficial similarity signals, including lexical similarity (\textit{Lexsim})~\cite{ratcliff1988pattern}, CodeBLEU~\cite{ren2020codebleu}, and embedding-based metrics from CodeBERT, GraphCodeBERT, UniXcoder, and BGE-Code~\cite{feng2020codebert, li2025towards}. We then employ Spearman's $\rho$~\cite{spearman1904} to assess the correlation between these similarity signals and model predictions. Finally, we construct a binary logistic regression model to quantify the explanatory power of these combined factors on the model's classification outcomes.



\begin{figure}[t]
    \centering
    \includegraphics[width=0.6\columnwidth]{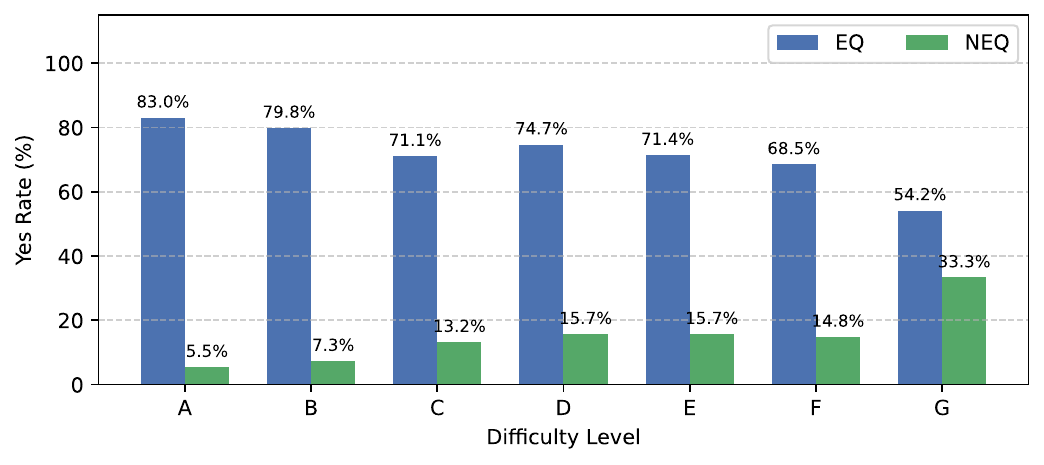}
    \caption{GPT-o4-mini's \textit{Yes} response rate across problem difficulty levels.}
    \label{fig:difficulty_AE}
\end{figure}

\noindent\textbf{Impact of Problem Difficulty.}
Figure~\ref{fig:difficulty_AE} shows the best-performing model's \textit{Yes} response rate across difficulty levels (A--G) for equivalent (\textit{Pass1 vs.\ Pass2}) and non-equivalent pairs (\textit{Pass1 vs.\ Fail}).
As difficulty increases, the \textit{Yes} rate drops from 83.0\% to 54.2\% for equivalent pairs, but rises from 5.5\% to 33.3\% for non-equivalent ones.
This indicates that higher difficulty degrades the model's discriminative power, increasing both false negatives and false positives, rather than merely making it more conservative.

\begin{tcolorbox}[enhanced, boxsep=0pt, before skip=5pt, after skip=5pt]
\textbf{Finding 4:} For GPT-o4-mini, problem difficulty reduces the model's ability to distinguish equivalent from non-equivalent pairs: as difficulty increases, the model misses more true equivalences while also producing more false equivalence judgments.
\end{tcolorbox}

\begin{figure}[ht!]
\centering
\begin{subfigure}[t]{0.49\textwidth}
    \centering
\includegraphics[width=\linewidth]{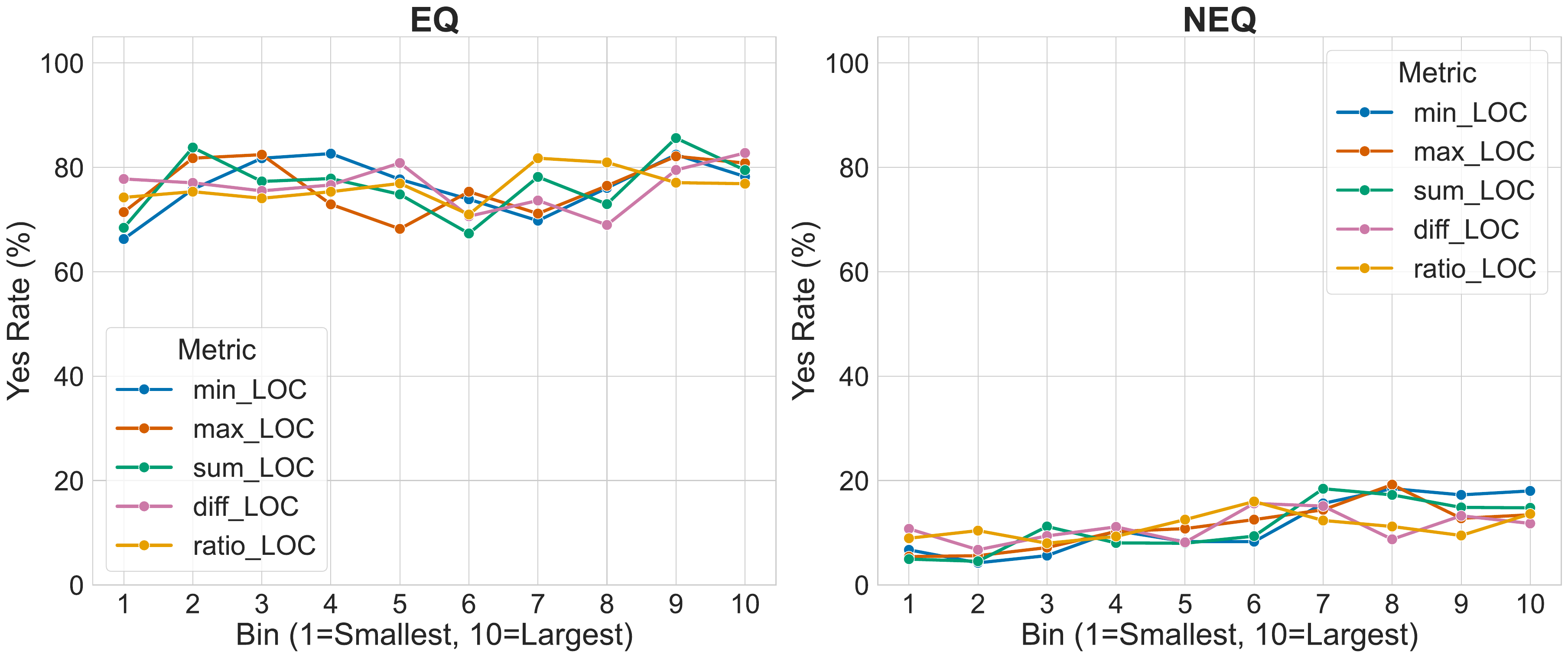}
    \caption{Code length}
    \label{fig:loc_overall}
\end{subfigure}
\hfill
\begin{subfigure}[t]{0.5\textwidth}
    \centering
    \includegraphics[width=\linewidth]{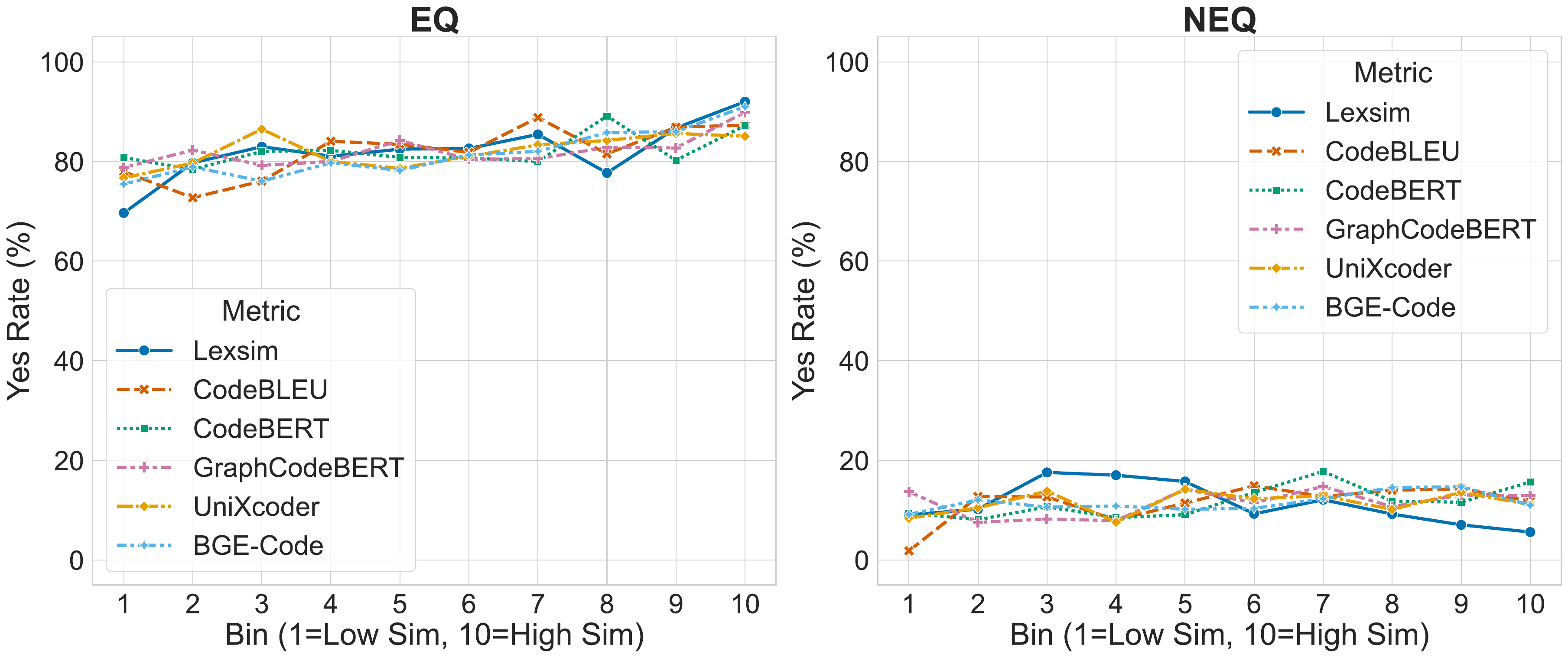}
    \caption{Code similarity}
    \label{fig:similarity_overall}
\end{subfigure}

\caption{GPT-o4-mini's \textit{Yes} response rate across code length and similarity bins. Left panels report equivalent pairs (EQ), and right panels report non-equivalent pairs (NEQ).}
\label{fig:loc_similarity_overall}
\end{figure}

\noindent\textbf{Impact of Code Length.} Figure~\ref{fig:loc_overall} shows GPT-o4-mini's \textit{Yes} response rate across LOC-based complexity bins. We consider five LOC-related measures: minimum LOC, maximum LOC, summed LOC, absolute LOC difference, and LOC ratio between the two programs. These measures allow us to distinguish the effect of overall input size from the effect of size imbalance between paired implementations.

Overall, we observed that code length affects NEQ pairs more clearly than EQ pairs. For EQ pairs, we observe high and stable \textit{Yes} rates across LOC bins, suggesting that larger code pairs do not substantially hinder equivalence recognition. For NEQ pairs, however, we observe that the \textit{Yes} rate increases in larger bins, indicating more false equivalence judgments as input size grows. This trend is stronger for \textit{min\_LOC}, \textit{max\_LOC}, and \textit{sum\_LOC} than for \textit{diff\_LOC} and \textit{ratio\_LOC}, suggesting that input size matters more than size imbalance.

\begin{tcolorbox} [enhanced, boxsep=0pt, before skip=5pt, after skip=5pt]
\textbf{Finding 5:} Code length mainly affects non-equivalent pairs: as overall input size increases, the model becomes more likely to produce false equivalence judgments.
\end{tcolorbox}

\noindent\textbf{Impact of Similarity Signals.} Figure~\ref{fig:similarity_overall} illustrates the best-performing model's \textit{Yes} response rate across similarity bins for lexical (Lexsim), structural (CodeBLEU), and embedding-based metrics (CodeBERT, GraphCodeBERT, UniXcoder, BGE-Code).

Similarity signals strongly correlate with correct \textit{Yes} judgments for equivalent (EQ) pairs, led by CodeBLEU ($\rho=0.87, p=0.001$) and UniXcoder ($\rho=0.84, p=0.002$), followed by BGE-Code, Lexsim, and GraphCodeBERT (all $p<0.05$).
For non-equivalent (NEQ) pairs, although overall \textit{Yes} rates remain low, higher similarity induces a shortcut-learning bias that increases false equivalences.
This bias is driven by structural and semantic features like CodeBERT ($\rho=0.76, p=0.011$) and CodeBLEU ($\rho=0.70, p=0.025$), whereas textual overlap (Lexsim) shows no significant impact ($p=0.138$).
Thus, similarity acts as a double-edged sword: aiding EQ detection while masking subtle differences in NEQ pairs.

\begin{tcolorbox}[enhanced, boxsep=0pt, before skip=5pt, after skip=5pt]
\textbf{Finding 6:} For GPT-o4-mini, similarity is associated with correctly identifying equivalent pairs, especially for CodeBLEU and UniXcoder. However, in NEQ pairs, CodeBERT and CodeBLEU are associated with more false equivalence predictions, indicating a residual similarity bias.

\end{tcolorbox}

\noindent\textbf{Regression Analysis.} We model GPT-o4-mini's \textit{Yes} predictions using logistic regression with problem difficulty, \textit{sum\_LOC}, CodeBLEU, programming language, ground-truth label (EQ vs.\ NEQ), and a difficulty\,$\times$\,label interaction as predictors. We use \textit{sum\_LOC} as the representative size metric because alternative LOC measures (min, max, and sum) show consistent trends. Although \textit{sum\_LOC} and difficulty are moderately correlated across languages and implementation states (Spearman’s $\rho \in [0.36,0.42]$,  Kendall’s $\tau \in [0.29,0.33]$), the association is not strong enough to indicate severe collinearity, supporting their inclusion as distinct predictors. Table~\ref{tab:regression_results} shows that the included factors account for substantial variation in the model's responses (Pseudo $R^2=0.429$; $n=2{,}969$ of 3{,}000 nominal instances, after excluding 5 problems with missing difficulty labels and 1 unparsable response).


\begin{table}[htbp]
\centering
\scriptsize
\caption{Logistic regression of factors associated with the model's \textit{True} predictions.}
\label{tab:regression_results}
\begin{tabular}{lrrrr}
\toprule
\textbf{Predictor} & $\beta$ & \textbf{SE} & \textbf{$p$-value} & \textbf{Odds Ratio} \\
\midrule
Intercept & 1.791 & 0.110 & $<0.001^{***}$ & 5.993 \\
Difficulty & -0.350 & 0.067 & $<0.001^{***}$ & 0.705 \\
Difficulty $\times$ NEQ & 0.804 & 0.103 & $<0.001^{***}$ & 2.234 \\
NEQ (vs. EQ) & -3.755 & 0.115 & $<0.001^{***}$ & 0.023 \\
Java (vs. CPP) & -0.183 & 0.145 & 0.208 & 0.833 \\
Python (vs. CPP) & -0.450 & 0.134 & $0.001^{**}$ & 0.637 \\
CodeBLEU & 0.212 & 0.053 & $<0.001^{***}$ & 1.236 \\
sum\_LOC & 0.056 & 0.060 & 0.353 & 1.058 \\
\midrule
\multicolumn{5}{l}{Observations: 2969 \quad Pseudo $R^2$: 0.429 \quad $^{**}p<0.01$ \quad $^{***}p<0.001$} \\
\bottomrule
\end{tabular}%
\end{table}

As expected, the ground-truth label is the dominant predictor: NEQ pairs are much less likely to receive a \textit{Yes} prediction than EQ pairs (OR $=0.023$, $p<0.001$), indicating that the model discriminates substantively. However, difficulty has a conditional effect. For EQ pairs, higher difficulty reduces the likelihood of a \textit{Yes} prediction ($\beta=-0.350$, OR $=0.705$, $p<0.001$), suggesting more false non-equivalence judgments. For NEQ pairs, the positive difficulty\,$\times$\,NEQ interaction ($\beta=0.804$, OR $=2.234$, $p<0.001$) indicates the opposite tendency: harder non-equivalent cases are more likely to be incorrectly accepted as equivalent. Similarity remains an independent predictor even after controlling for label, difficulty, language, and code length. CodeBLEU significantly increases the odds of a \textit{Yes} prediction ($\beta=0.212$, OR $=1.236$, $p<0.001$), suggesting a residual reliance on surface or structural resemblance. In contrast, \textit{sum\_LOC} is not significant ($p=0.353$), indicating that code length does not independently explain the model's decisions once the other factors are controlled. Language effects are also limited but not absent: Java does not differ significantly from CPP ($p=0.208$), whereas Python is associated with fewer \textit{Yes} predictions ($\beta=-0.450$, OR $=0.637$, $p=0.001$), suggesting more conservative behavior for Python pairs.

Overall, the regression results refine the previous finding. Code similarity matters, but it is not the only driver of LLM equivalence judgments. The model's decisions are shaped by the interaction between ground-truth equivalence and difficulty, residual similarity effects, and language-specific behavior, while code size has no significant independent effect.

\begin{tcolorbox}[enhanced, boxsep=0pt, before skip=5pt, after skip=0pt]
\textbf{Finding 7:} For GPT-o4-mini, equivalence judgments are not explained by similarity alone: logistic regression shows that its predictions are mainly shaped by the ground-truth label, difficulty conditioned on equivalence status, CodeBLEU, and Python-specific conservativeness, while code length has no significant independent effect.
\end{tcolorbox}

\subsection{Why LLMs Fail to Judge Code Functional Equivalence (\texorpdfstring{RQ$_4$}{RQ4})}

To answer \textbf{RQ$_4$}, we analyzed misclassified code pairs from GPT-o4-mini to uncover the root causes of LLM failures. Rather than relying solely on aggregate metrics, we investigated whether errors stemmed from systematic prediction biases, superficial similarities, language-specific misunderstandings, abstraction mismatches, or stochastic instability. 

Given PolyHuman's scale, we adopted an incremental evaluation strategy guided by thematic saturation~\cite{glaser2017discovery, guest2006many}. Each pair was evaluated three times using majority voting, and we terminated manual error classification when 16 consecutive erroneous instances yielded no new error categories. In total, we evaluated 900 instances across 60 problems. GPT-o4-mini exhibited an instability rate of 18\% (162 instances with inconsistent predictions across runs). For the manual thematic analysis, we focused on systematic errors (i.e., instances where the model failed in at least two of the three runs) and leveraged a Chain-of-Thought (CoT) prompt (presented below) to elicit and analyze the model's underlying reasoning.

\begin{tcolorbox}[
  title={Prompt used for RQ$_4$ error analysis},
  label={prompt:rq4-cot},
  breakable,
  colback=gray!3,
  colframe=gray!95,
  fonttitle=\bfseries\small,
  boxrule=0.5pt,
  arc=1pt,
  left=4pt,
  right=4pt,
  top=0pt,
  bottom=0pt
]
\scriptsize
\ttfamily
\raggedright
\noindent
We define functionally equivalent as: two code snippets produce the same output for every possible input.
They may be written in different programming languages; ignore stylistic or structural differences and focus only on input--output behaviour.

\medskip
\noindent
Your task: determine whether \texttt{<code1>} and \texttt{<code2>}, which are both intended to solve the problem described in \texttt{<Problem>}, are functionally equivalent.

\medskip
\noindent
\texttt{<Problem>} \{problem\} \texttt{</Problem>}\\
\texttt{<code1>} \{code1\_numbered\} \texttt{</code1>}\\
\texttt{<code2>} \{code2\_numbered\} \texttt{</code2>}

\medskip
\noindent
Think step-by-step before giving your final answer.

\medskip
\noindent
Respond only with a JSON object using the following schema:

\begin{lstlisting}[basicstyle=\ttfamily\scriptsize, breaklines=true, columns=fullflexible, frame=none,
xleftmargin=0pt, 
xrightmargin=0pt,
aboveskip=0pt,
belowskip=0pt,
resetmargins=true,
backgroundcolor=\color{gray!3}]
{
  "verdict": "Yes" | "No",
  "reasoning": "<step-by-step explanation>",
  "key_lines": {
    "code1": [<integer line numbers>],
    "code2": [<integer line numbers>]
  }
}
\end{lstlisting}
\end{tcolorbox}

From the 900 evaluated instances, we identified 81 cases that met the systematic disagreement criterion and manually analyzed them. Table~\ref{tab:highlight_pairs} summarizes their distribution by pair type, covering same-language and cross-language pairs with equal or different outcomes. When considering subtask opportunity bounds (360 equivalent vs.\ 540 non-equivalent pairs), false negatives occurred at a slightly higher rate than false positives ($9.72\%$ vs.\ $8.52\%$, corresponding to 35 FNs and 46 FPs). In terms of sample composition, cross-language comparisons accounted for 46 of the 81 analyzed failure instances.

\begin{table*}[ht!]
\centering
\caption{Pair-type distribution of the 81 manually analyzed instances.}
\vspace{-0.3cm}
\label{tab:highlight_pairs}
\tiny
\setlength{\tabcolsep}{3.2pt}
\renewcommand{\arraystretch}{1.05}
\resizebox{\textwidth}{!}{%
\begin{tabular}{lr@{\hspace{0.8em}}lr@{\hspace{0.8em}}lr@{\hspace{0.8em}}lr}
\toprule
\multicolumn{2}{c}{\textbf{\textit{Same language, equal outcome}}} &
\multicolumn{2}{c}{\textbf{\textit{Same language, different outcome}}} &
\multicolumn{2}{c}{\textbf{\textit{Different language, equal outcome}}} &
\multicolumn{2}{c}{\textbf{\textit{Different language, different outcome}}} \\
\cmidrule(lr){1-2}
\cmidrule(lr){3-4}
\cmidrule(lr){5-6}
\cmidrule(lr){7-8}
\textbf{Pair} & \textbf{N} &
\textbf{Pair} & \textbf{N} &
\textbf{Pair} & \textbf{N} &
\textbf{Pair} & \textbf{N} \\
\midrule

Java Pass1 vs. Java Pass2     & 9 &
Java Pass1 vs. Java Fail      & 8 &
CPP Pass1 vs. Java Pass1      & 6 &
Java Fail vs. Python Pass1    & 9 \\

Python Pass1 vs. Python Pass2 & 4 &
CPP Pass1 vs. CPP Fail        & 6 &
Python Pass1 vs. CPP Pass1    & 5 &
CPP Fail vs. Java Pass1       & 7 \\

CPP Pass1 vs. CPP Pass2       & 6 &
Python Pass1 vs. Python Fail  & 2 &
Java Pass1 vs. Python Pass1   & 5 &
Java Fail vs. CPP Pass1       & 6 \\

                              &   &
                              &   &
                              &   &
CPP Fail vs. Python Pass1     & 3 \\

                              &   &
                              &   &
                              &   &
Python Fail vs. Java Pass1    & 3 \\

                              &   &
                              &   &
                              &   &
Python Fail vs. CPP Pass1     & 2 \\

\bottomrule
\end{tabular}
}
\vspace{-0.3cm}
\end{table*}

To examine whether these failures were specific to \mbox{GPT-o4-mini} or reflected broader limitations of current LLMs, we evaluated two additional models on the same 81 instances using the same CoT prompt. We found that 24 instances were misclassified by all three models, suggesting intrinsically difficult cases. However, several \mbox{GPT-o4-mini} failures were correctly resolved by Claude-Opus-4.7, Gemini-3-Flash, or both (19 resolved by both), indicating model-specific differences. In addition, nine \mbox{GPT-o4-mini} failures were corrected after applying CoT prompting. We consider a misclassification corrected when CoT prompting produces predictions consistent with the ground truth across three independent runs.

\begin{tcolorbox}[enhanced, boxsep=0pt, before skip=5pt, after skip=5pt]
\textbf{Finding 8:}
Among the three models evaluated on the 81 disagreements, failures reflect both shared semantic reasoning limitations and model-specific weaknesses. While 24 instances were missed by all models, other GPT-o4-mini failures were correctly resolved by Claude-Opus-4.7, Gemini-3-Flash, or after applying CoT prompting. This suggests that some failures reflect insufficient reasoning depth rather than a complete lack of capability. However, CoT alone was not enough to ensure robust equivalence detection.
\end{tcolorbox}

\textbf{Qualitative analysis of failure modes.}
Overall, our analysis pipeline filter proceeded as follows: $900$ initial instances $\rightarrow$ $162$ unstable instances $\rightarrow$ $81$ systematic disagreements $\rightarrow$ $9$ corrected by CoT $\rightarrow$ $72$ coded failure cases ($55$ genuine model failures and $17$ ground-truth issues).
To better understand why LLMs fail, we manually inspected the remaining 72 misclassified cases and identified recurrent (sub-)categories of failure modes. We organize these 72 cases into three top-level categories: \textit{Knowledge Failure} (part of the 55 model failures), \textit{Abstraction-Level Reasoning Failure} (part of the 55 model failures), and \textit{LLM Judgment Is Actually Correct} (the 17 ground-truth issues). These categories capture different mechanisms behind apparent failures: incorrect knowledge assumptions, breakdowns across abstraction levels, and defensible disagreements with the ground truth. The complete qualitative analysis, including the full set of coded instances and examples, is available in our replication package~\cite{sun2026esem}.

We can observe that most cases fall under \textit{Abstraction-Level Reasoning Failure} (38/72, 52.8\%), especially \textit{Implementation-Level Detail Error} (21/72, 29.2\%). This suggests that LLMs often produce plausible explanations but fail to verify concrete details such as index correspondence, state updates, control-flow paths, or variable evolution. We describe each category and sub-category in detail as follows: figures show only the relevant code evidence and failure mechanisms, while LLM claims are discussed in the text.

\begin{figure}[ht!]
\centering
\begin{subfigure}[t]{0.49\textwidth}
    \centering
\includegraphics[width=\linewidth]{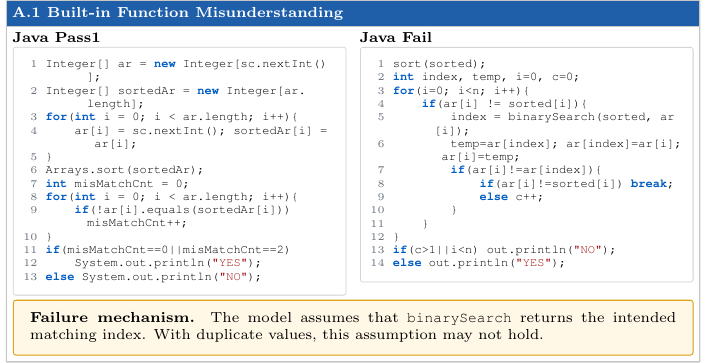}
    \label{fig:A1}
\end{subfigure}
\begin{subfigure}[t]{0.502\textwidth}
    \centering
    \includegraphics[width=\linewidth]{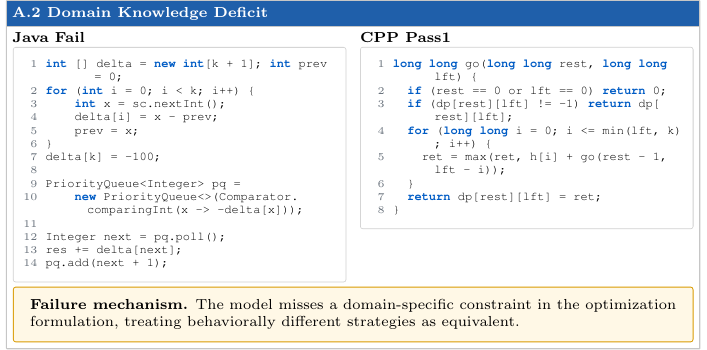}
    \label{fig:A2}
\end{subfigure}
\vspace{-0.5cm}
\caption{Representative examples of knowledge failures.}
\label{fig:A1A2}
\end{figure}

\noindent\textbf{A. Knowledge Failure (17).}
This category captures cases in which the LLM lacks or misapplies factual knowledge relevant to judging code functional equivalence, causing its reasoning to rest on incorrect foundations. Figure~\ref{fig:A1A2} illustrates the representative examples.

\begin{itemize}
  \item \textbf{A.1 Language/Built-in Function Misunderstanding (11)} occurs when the LLM misinterprets the behavior of a programming language construct or built-in function. In \textit{FP-row-19}, shown in Figure~\ref{fig:A1A2}(A.1), the LLM claims that both snippets decide whether an array can be sorted with at most one swap. However, its reasoning depends on the call to \texttt{binarySearch} in line~3. When duplicate elements are present, this call may return any matching index, not necessarily the index assumed by the model. As a result, the subsequent swap logic in lines~4--6 may not preserve the intended behavior.
  \item \textbf{A.2 Domain Knowledge Deficit (5)} captures cases where the model lacks the mathematical, algorithmic, or problem-specific knowledge to evaluate equivalence. In \textit{FP-row-32}, shown in Figure~\ref{fig:A1A2}(A.2), the LLM treats two optimization strategies as equivalent. However, the excerpt highlights a domain-specific adjustment in line~8, where \texttt{delta[k]} is set to \texttt{-100}. This value changes the optimization behavior, but the model overlooks its role and incorrectly assumes that the two strategies compute the same contribution.

  \item \textbf{A.3 Equivalence Criterion Confusion (1)} occurs when the LLM misinterprets what ``functional equivalence'' means in the context of this task. For example, it may treat passing all test cases on an online judge as sufficient evidence of functional equivalence or conflate semantic equivalence with functional equivalence.
\end{itemize}

\begin{figure}[ht!]
\centering
\begin{subfigure}[t]{0.32\textwidth}
    \centering
\includegraphics[width=\linewidth]{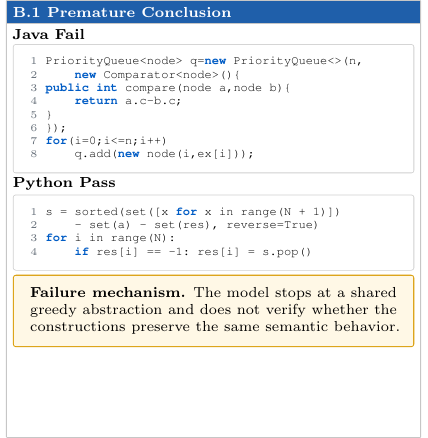}
    \label{fig:B1}
\end{subfigure}
\begin{subfigure}[t]{0.32\textwidth}
    \centering
    \includegraphics[width=\linewidth]{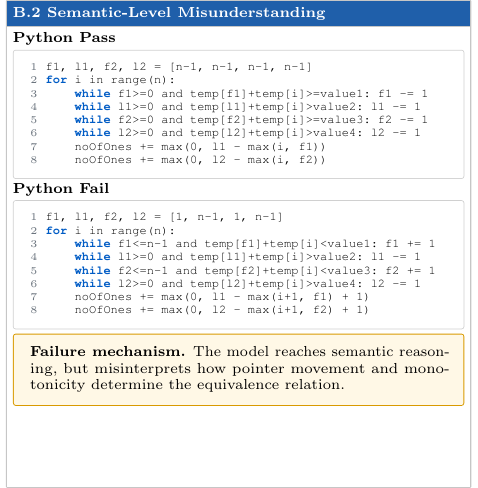}
    \label{fig:B2}
\end{subfigure}
\begin{subfigure}[t]{0.32\textwidth}
    \centering
\includegraphics[width=\linewidth]{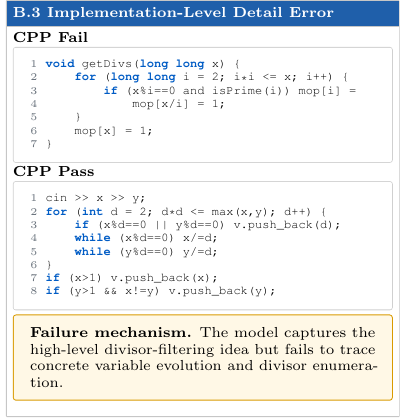}
    \label{fig:B3}
\end{subfigure}
\vspace{-0.5cm}
\caption{Representative examples of abstraction-level reasoning failure.}
\label{fig:B-examples}
\end{figure}

\noindent\textbf{B. Abstraction-Level Reasoning Failure (38).} This category captures cases in which the LLM has relevant knowledge but fails while moving across levels of abstraction. The model may start from a plausible algorithmic interpretation, but the reasoning chain breaks before it correctly verifies the actual behavior of the two solutions. These failures are therefore not simply about whether the model recognizes the algorithm, but whether it can connect that recognition to a correct semantic and execution-level judgment. Figure~\ref{fig:B-examples} illustrates the representative examples.

\begin{itemize}
  \item \textbf{B.1 Premature Conclusion (Pre-Semantic) (11)} occurs when the LLM stops at surface-level or algorithm-level similarity and concludes equivalence before performing a semantic comparison. In \textit{FP-row-18}, shown in Figure~\ref{fig:B-examples}(B.1), the LLM claims that both snippets implement the same greedy construction for preserving prefix MEX behavior. This claim is based on recognizing a priority-queue based selection in lines~1--7 of the first snippet and a set-based filling strategy in lines~1--3 of the second snippet. However, the model does not verify whether these two constructions preserve the same input-output behavior. The failure, therefore, occurs before a semantic comparison is performed.

  \item \textbf{B.2 Semantic-Level Misunderstanding (6)} occurs when the model reaches the semantic level but misinterprets the key behavioral property that determines equivalence. In \textit{FP-row-13}, shown in Figure~\ref{fig:B-examples}(B.2), the model correctly identifies pointer movement as relevant. The critical evidence appears in lines~3--6 of both snippets: in the first snippet, pointers such as \texttt{f1} and \texttt{f2} move left, whereas in the second snippet they move right. The LLM interprets this directional difference as evidence of non-equivalence. However, its conclusion is based on an incorrect understanding of the monotonicity invariant governing the pair-counting logic. Thus, unlike B.1, the model does reason semantically, but the semantic interpretation is flawed.
 
\item \textbf{B.3 Implementation-Level Detail Error (21)} occurs when the model captures the high-level intent but fails to verify concrete execution details. In \textit{FP-row-6}, Figure~\ref{fig:B-examples}(B.3), the LLM claims that both snippets enumerate candidate prime divisors and filter them across pairs. The key issue is in the concrete divisor extraction logic. In the first snippet, lines~2--4 add factors based on divisibility but do not update \texttt{x} by removing the factors already found. In contrast, the second snippet explicitly divides \texttt{x} and \texttt{y} inside the loops in lines~4--5, exposing remaining factors that may be added later in lines~7--8. The LLM overlooks this variable evolution, leading to an incorrect equivalence judgment.

\end{itemize}

\noindent\textbf{C. LLM Judgment Is Actually Correct (17).} This category captures cases in which the LLM's judgment is in fact correct; the disagreement between ground-truth and GPT-o4-mini prediction lies in the data construction or validation rather than in the LLM's reasoning. This category encompasses two sub-cases: either a counterexample exists that contradicts the ground-truth label, confirmed via manual execution of test cases; or the problem admits multiple valid outputs, such that the two code snippets are genuinely not functionally equivalent, representing an internal threat rather than an LLM's reasoning failure.




\begin{itemize}

  \item \textbf{C.1 LLM Identifies a Genuine Counterexample (4)} refers to cases where the LLM successfully identifies a test case that produces different outputs from the two code snippets, or correctly recognizes a potential runtime error scenario (e.g., behavioral divergence caused by hash collisions).

  \item \textbf{C.2 Problem Has More Than One Valid Solution (13)} refers to cases where the original problem inherently admits multiple correct implementations. The LLM identifies an alternative valid solution beyond the reference answer, making its ``not equivalent'' judgment logically defensible.

\end{itemize}

\begin{tcolorbox}[enhanced, boxsep=0pt, before skip=5pt, after skip=0pt]
\textbf{Finding 9:} Among the GPT-o4-mini errors analyzed qualitatively, the primary bottleneck is an inability to accurately trace microscopic dynamic execution paths and sequential variable evolution. Of the two additional models evaluated on these cases, Opus alone resolved 17 instances (14 Abstraction-Level Reasoning Failure) and Gemini alone resolved 12 (6 Knowledge Failure), suggesting Opus is comparatively stronger at abstraction-level reasoning and Gemini at knowledge-related cases.
Additionally, we identified at least $17$ ground-truth labeling flaws among the analyzed failure cases ($1.89\%$ of the total $900$ instances).\footnotemark
\end{tcolorbox}
\footnotetext{Note that this represents a lower bound on dataset label noise, as manual inspection was restricted to targeted, non-random failure instances where the model disagreed with the ground truth.}
\section{Actionable Implications}

Our analysis of GPT-o4-mini shows that the model does not rely solely on surface-level code similarity when judging functional equivalence; instead, its predictions are shaped by a structured combination of factors operating at different levels of abstraction (quantitatively, these factors yield a Pseudo $R^2 = 0.429$, with ground-truth equivalence dominating over code similarity). However, these very factors also give rise to systematic failure patterns that manifest at distinct levels of abstraction. Based on these findings, we derive several actionable implications for researchers and practitioners.

\noindent\textbf{Implications for LLM Researchers.}
Our findings reveal that GPT-o4-mini's failures stem primarily from deficient implementation-level comprehension and fragile domain knowledge, rather than mere surface similarity. However, surface-matching shortcuts persist: even when controlling for difficulty and ground-truth labels, GPT-o4-mini relies on code similarity to a small but statistically robust degree ($\beta = 0.212$, $p < 0.001$), suggesting that future work must prioritize fine-grained semantic modeling over generalized abstract reasoning. Crucially, the model breaks down consistently on three kinds of inputs: (1) tangled control flow (loops + branching), (2) structurally diverse human-written code, and (3) cross-language pairs, where a language-specific bias is evident as the model exhibits pronounced caution exclusively on Python ($\beta = -0.450$, $\text{OR} = 0.637$, $p = 0.001$) while CPP and Java remain indistinguishable ($p = 0.208$). Furthermore, average accuracy masks a severe asymmetric vulnerability driven by a powerful difficulty$\times$equivalence interaction ($\beta = 0.804$, $p < 0.001$), meaning that on harder problems the model increasingly misclassifies non-equivalent code as equivalent. Therefore, future benchmarks should supplement aggregate accuracy with stratified metrics to prevent these asymmetric failures from remaining invisible.

\noindent\textbf{Implications for Agent and Multi-Agent System Researchers.}
When LLMs (GPT-o4-mini, Claude-Opus-4.7, Gemini-3-Flash) encounter specific layer gaps, their capacity for self-correction is substantially diminished, suggesting that human-in-the-loop validation may be valuable in high-risk settings~\cite{mehtiyev2026beyond}.
Beyond individual agents, our analysis reveals complementary capability profiles across models: Claude-Opus-4.7 performed better on abstraction and intermediate reasoning, while Gemini-3-Flash shows a comparative advantage on cases requiring domain- or language-specific factual knowledge. Thus, multi-agent system designers can exploit this complementarity by assigning reasoning-intensive stages to the former and knowledge verification to the latter, enabling more effective model specialization.

\noindent\textbf{Implications for LLM-Assisted Programming Practitioners.}
Practitioners should exercise heightened scrutiny when deploying models for code involving complex loops, iterations, and multi-layered conditional logic, as model reliability degrades significantly under these constructs. Furthermore, cross-run analysis indicates that isolated correct outputs reflect inconsistent capability rather than robust competence.
\section{Threats to Validity}

\noindent\textbf{Internal Validity.}
PolyHuman labels two solutions passing all online judge test cases as functionally equivalent. However, test suites may not achieve full input coverage, and passing programs may still differ on unseen edge-case inputs. This threat is evidenced by Category C in our manual analysis, which accounts for 17 out of 900 instances (1.9\%), though its overall impact remains limited. A further threat is data contamination: CodeContests is public and likely present in the training data of the evaluated models, so if anything this would bias results toward overestimating model performance, making our finding of poor functional-equivalence judgment a conservative estimate.

\noindent\textbf{External Validity.}
PolyHuman is constructed from competitive programming platforms, where code tends to be algorithm-intensive and performance-oriented, which may not fully generalize to real-world code. Regarding model selection, our evaluation is primarily limited to open-source models in the 3B--20B parameter range, plus GPT-o4-mini due to resource constraints. However, the additional evaluation of Claude-Opus-4.7 and Gemini-3-Flash on the 81 manually analyzed instances in RQ4 partially mitigates this threat.

\noindent\textbf{Construct Validity.}
Although accuracy is used as the primary metric for cross-dataset comparability, we mitigate its limitations by separately analyzing Pass vs. Pass and Pass vs. Fail pairs, which is equivalent to reporting True Positive Rate and True Negative Rate independently. This decomposition provides a more informative view of model behavior under prediction bias than aggregate accuracy alone.

\noindent\textbf{Conclusion Validity.}
The manual classification of 81 failure cases into categories A, B, and C involves inherent subjectivity, as category boundaries depend on researcher interpretation. To mitigate this threat, all classifications were independently verified by the second author, disagreements resolved through discussion. We acknowledge that no formal inter-rater reliability measure (e.g., Cohen's Kappa) was computed, which limits the reproducibility of categorical assignments. All coded instances are made available in our replication package. 

\section{Related Work}

\noindent\textbf{LLMs for Code Understanding.} LLMs have achieved strong performance across code generation, summarization, and translation~\cite{jiang2026survey}, raising expectations that they can reason about program semantics like human developers. Yet a growing line of empirical SE research finds their performance highly inconsistent across tasks and settings~\cite{liu2024exploring, schafer2023empirical, Molison2025ESEM}. Our work extends this critical evaluation to a capability not yet examined under realistic conditions: judging code functional equivalence.

\noindent\textbf{Functional Equivalence and Clone Detection.} Functional equivalence underpins refactoring, migration, and clone detection~\cite{cheng2025codemenv, fowler2018refactoring, inoue2024improving, roy2009comparison}. Two recent benchmarks for LLMs target those tasks: EquiBench~\cite{wei2025equibench} and SeqCoBench~\cite{maveli2025can}. While valuable, both construct non-equivalent pairs via controlled transformations (OJ-V for EquiBench), 
so the equivalent/non-equivalent distinction is often local and can be solved by similarity shortcuts rather than semantic reasoning. This mirrors the similarity-driven tendency in clone detection~\cite{roy2007survey, svajlenko2014towards}, meaning high benchmark accuracy may not reflect true semantic understanding.

\noindent\textbf{Cross-Language Code Analysis.} Similarity shortcuts fail most clearly across languages, where equivalent programs differ drastically in syntax and idiom~\cite{moumoula2025struggles, perez2019cross}. LLMs are known to introduce bugs in cross-language translation~\cite{pan2024lost} and struggle with cross-lingual clone detection~\cite{moumoula2025struggles}, but these efforts address clone detection or translation rather than strict functional equivalence, and rely on synthetic data. Whether LLMs can reliably judge cross-language functional equivalence on human-written code remains open.

\noindent\textbf{Positioning.} We differ from prior work in three ways: (1) PolyHuman uses independently written human solutions rather than synthetic pairs; (2) we conduct the first systematic evaluation of cross-language functional equivalence judgment; and (3) beyond aggregate accuracy, we provide a mechanistic failure analysis across abstraction levels and discuss complementary-model strategies~\cite{zhou2025se}.

\section{Conclusion}
This paper investigates LLMs' ability to judge functional equivalence across programming languages. To address the issue that single-language and synthetic benchmarks overestimate LLMs' semantic understanding, we introduce PolyHuman, a novel dataset of human-written programs in CPP, Java, and Python. 

Our main contributions are twofold. Quantitatively, we observe a sharp decline in performance on human-written code. We find that decision biases appear more strongly model-dependent than uniformly language-driven, suggesting that model selection and ensembling are more critical than language choice. Furthermore, while superficial similarity exerts a secondary influence, errors are highly difficulty-conditional: on complex problems, GPT-o4-mini increasingly misclassifies non-equivalent code as equivalent.
Qualitatively, we identify two major classes of model failures, namely knowledge failures and abstraction-level reasoning failures, with the latter spanning presemantic, semantic, and implementation-level reasoning.
GPT-o4-mini demonstrates sound reasoning across many problems, albeit with some instability. 
We also identified 17 ground-truth/data issues among the analyzed cases; because these cases were selected based on model disagreement, this proportion should not be interpreted as an estimate of the dataset’s overall label-error rate.

Future work should investigate strategies to improve LLM semantic reasoning for code functional equivalence, particularly in cross-language and human-written scenarios. Promising directions include developing benchmarks with more diverse real-world programs, reducing reliance on superficial similarity signals, and improving reasoning stability across runs. Future studies may also explore ensemble and multi-agent approaches that combine complementary model strengths, as well as techniques for better tracing of execution paths and variable evolution in complex control-flow structures. 



\bibliographystyle{plainurl} 
\bibliography{references}

\end{document}